\documentclass[12pt,letterpaper]{article}

\usepackage{amsmath,amssymb,calc}
\usepackage{graphicx}\usepackage{pdfsync}
\usepackage{float}
\usepackage{amstext}    
\usepackage{array}    
\usepackage{color}

\newcommand\be{\begin{equation}}
\newcommand\ee{\end{equation}}
\newcommand{\bea}{\begin{eqnarray}}
\newcommand{\eea}{\end{eqnarray}}

\newcommand{\nn}{\nonumber}

\def\id{\protect{{1 \kern-.28em {\rm l}}}}

\def\1{^{(1)}}
\def\0{^{(0)}}
\def\2{^{(2)}}

\def\id{\protect{{1 \kern-.28em {\rm l}}}}

\let\non\nonumber

\begin{document}
\begin{titlepage}
\begin{center}
\hfill \\

\vskip 2cm

{\Large \bf Axion Stars in the Infrared Limit}
\vskip 1.5 cm
{\it  Joshua Eby, Peter Suranyi, Cenalo Vaz, and L.C.R. Wijewardhana
\\
   Dept. $\!$of Physics, University of Cincinnati,
Cincinnati, OH 45221, USA}\non\\

\begin{abstract}

Following Ruffini and Bonazzola, we use a quantized boson field to describe condensates of axions forming compact objects.  Without 
substantial modifications, the method can only be applied to axions with decay constant, $f_a$, satisfying $\delta=(f_a\,/\,M_P)^2\ll 1$, 
where $M_P$ is the Planck mass. Similarly, the applicability of the Ruffini-Bonazzola method to axion stars also requires that the 
relative binding energy of axions satisfies $\Delta=\sqrt{1-( E_a\,/\,m_a)^2}\ll1$, where $E_a$ and $m_a$ are the energy and mass of 
the axion.  The simultaneous expansion of the equations of motion in $\delta$ and $\Delta$ leads to a simplified set of equations, 
depending only on the parameter, $\lambda=\sqrt{\delta}\,/\,\Delta$ in leading order of the expansions. Keeping leading 
order in $\Delta$ is equivalent to the infrared limit, in which only relevant and marginal terms contribute to the equations of 
motion.  The number of axions in the star is uniquely determined by $\lambda$. Numerical solutions are found in a wide range of 
$\lambda$.  At small $\lambda$ the  mass and radius of the axion star rise linearly with $\lambda$. While  at larger $\lambda$ the 
radius of the star continues to rise, the mass of the star, $M$, attains a maximum at $\lambda_{\rm max}\simeq 0.58$. All stars 
are unstable for $\lambda>\lambda_{\rm max}$. We discuss the relationship of our results to current observational constraints 
on dark matter and the phenomenology of Fast Radio Bursts.
\end{abstract}

\hfill \\
\end{center}
\vskip 2 cm

\end{titlepage}

\section{Introduction}
Scalar fields, which give rise to spin-zero quanta satisfying Bose statistics, are natural in quantum field theories of phenomenological 
importance. The recently discovered Higgs field, which generates masses for all the other particles,  is the prime example.  The axion, 
a yet to be discovered pseudoscalar, postulated to solve the strong-CP problem endemic to QCD, is another well motivated spin-zero 
particle. It is also generic in string theory for  4-dimensional axion-like degrees of freedom to arise as Kaluza-Klein zero modes from 
the  compactification of anti-symmetric tensor fields defined in 10 space-time dimensions. Quintessence, the almost massless scalar 
field invoked to drive the late-time acceleration of the universe, is another hypothetical scalar degree of freedom.  It would be quite 
interesting to determine if such spin-zero degrees of freedom could form stable compact structures due to their self-gravitation.

Starting with the seminal works of Kaup~\cite{Kaup} and Ruffini and Bonazzola~\cite{RB}, the study of gravitationally bound  bosonic  
degrees of freedom  has received a great deal of attention.    It has been seen that complex scalar field configurations in the 
presence of gravity can form stable compact objects, termed boson stars ~\cite{CSW}, ~\cite{SS}, ~\cite{bosonstars}.  Kaup in his original 
work solved the Klein-Gordon free field equation in an asymptotically flat  spherically symmetric  space-time background and found 
localized  spherically symmetric field configurations which are  energy eigenstates.  Adopting a slightly different approach, Ruffini 
and Bonazzola used the expectation value of the energy-momentum tensor operator of a second quantized free hermitian scalar field 
system, evaluated in an $N$-particle state where all the particles occupy the lowest energy state, to be the source of gravitational 
interactions. Here the basis states for second quantization were the wave functions of the Klein-Gordon equation. This approach yielded 
an equation of motion for the wave function of the $N$-particle state, which represented a condensate of $N$ bosons in a single state. 
They found stable localized wave functions for the condensate, indicating the interesting possibility that lumps of self-gravitating 
scalar field configurations could exist in nature. 

In a recent set of publications, ~\cite{Barranco}  ~\cite{Barranco2} analyzed the self-gravitating field theoretical system consisting 
of a real scalar field with an interaction potential of the form $ f_a{}^2 m_a{}^2 (1-\cos (\frac{\phi}{f_a})) $. They followed the 
method pioneered by Ruffini and Bonazzola. Such a potential represents the interactions of the axion field at low energies, as derived 
using the dilute instanton gas approximation, where $f_a$ represents the axion decay constant, and $m_a$ the axion 
mass~\cite{GrossPisarskiYaffee}. In this paper we analyze the same system but present an analytic expansion, which simplifies the 
equations of motion. We also consider a different range of input parameters. This leads to  interesting physical consequences. Without 
substantial modification, our expansion method can only be applied to axions with $f_a \ll M_P$, where $M_P=1/\sqrt{8\pi G}$ is the 
Planck mass.  In fact, as we will point out later, the field theoretic method of Ruffini and Bonazzola \cite{RB} can only be applied 
in this regime when the relative binding energy of axions, defined by $\Delta= \sqrt{m_a{}^2-E_a{}^2}\,/\,  m_a$, is small. We 
expand the Einstein and Klein-Gordon equations in terms of a series in $\Delta$. We find that the contribution of all operators
with  scaling dimension $d > 4$    vanish as a power of $\Delta$. The leading order equations depend on the single parameter 
$\lambda ={{ f_a} \over {M_{P}\Delta}} $. We find analytic formulas for the mass and the radius of the star as a function of 
$\lambda$, satisfied in the ranges $f_a\,/\,M_P \ll \lambda \ll M_P \,/\,f_a$.  Parametrizing solutions of the equations of motion by 
$\lambda$, rather than the central density of the star,  has the advantage of being able to compare total energies of solutions with 
equal numbers of axions as shown in Fig.   \ref{Mplot}. As expected, the critical temperature of the condensed axion star is very 
large, $T_c>10^{10}$ GeV.

\section{Axion field Dynamics in the Condensed State}
\setcounter{equation}{0}

The dynamics of a self-gravitating, free hermitian scalar field was first analyzed in \cite{RB}.  More recently, the authors of 
\cite{Barranco} used a similar procedure to describe the condensation of interacting hermitian Bose fields.  We revisit this analysis 
by evaluating the expectation value of the axion potential in the $N$ particle condensed state and simplifying the equations of motion 
using the expansion method described in the previous section.  As stated earlier the range of parameters we explore  differ from that 
of \cite{Barranco}, \cite{Barranco2}.

\subsection{Expectation Values}

In our analysis, we consider axion dynamics to be described by a hermitian scalar field $\phi_a$ with potential energy 
\be\label{potential1}
W(\phi_a)=f_a{}^2\,m_a{}^2\left[1-\cos\left(\frac{\phi_a}{f_a}\right)\right].
\ee
We will see that the leading-order deviation of the space-time metric from flat space is proportional to $\delta \equiv f_a^2/M_P^2$, 
the effective coupling constant of matter to gravity.\footnote {Our expansion parameter $\delta$ is related to parameter $\Lambda$, 
used in \cite{bosonstars} and \cite{Barranco} as $\delta=(24\,\pi\,\Lambda)^{-1}.$} If we therefore restrict ourselves to decay 
constants $f_a \lesssim 10^{-2} M_P$, it is sufficient to take gravity into account to leading order in $\delta \lesssim 10^{-4}$.

We start with the standard definition of the axion potential (\ref{potential1}), but replace it by the expectation value of a quantum 
potential in which the quantum field, expanded in modes having definite radial, polar and azimuthal quantum numbers, is
\be\label{phi1}
\phi_a\to \Phi(x)=\Phi^-(x)+\Phi^+(x)\equiv\sum_{n,l,m}R_{n,l}(r)\,[e^{i\,E_{n,l}t}Y_l^m(\theta,\phi)\,a_{n,l,m}+h.c.],
\ee
where $a^\dagger_{n,l,m}$ is the creation operator for an axion with the appropriate quantum numbers.  To a very good 
approximation every axion in an axion star is in the ground state, so the star is an almost perfect condensate.  As we will see 
later, for QCD axions the critical temperature of the condensate is $10^{11}$ GeV. Consequently, in what follows, we will 
use the expression
\be
\Phi(x)=R(r)[e^{i\,E\,t}a+e^{-i\,E\,t}a^\dagger]
\ee
and, in Appendix \ref{Appendix1}, we calculate the expectation value of the axion potential exactly in the tree 
approximation. Note that, to justify using the tree approximations one needs to consider the fact that negative energy states and 
with them loop corrections can be neglected if binding energies are of $\Delta E_a \ll m_a$. We obtain in the large $N$ limit
\be\label{potexp}
\langle N \vert \cos(\Phi\,/\,f_a)\vert N\rangle\simeq J_0(2\,\sqrt{N}\,R\,/\,f_a),
\ee
an effective potential that is different from the classical potential. We thus 
obtain the equations of motion by taking the expectation value of the Einstein equation and the scalar field equation. For this, 
we will also need the expectation values
\bea
\langle N \vert\Phi^2\vert N\rangle&\simeq&2\,N^2 R(r)^2,\nn\\
\langle N-1 \vert\Phi\vert N\rangle&\simeq&\sqrt{N} R(r)\,e^{-i\,E\,t}.
\eea

\subsection{Equations of motion}

We write the spherically symmetric metric as
\be
ds^2=B(r)\,dt^2-A(r)\,dr^2-r^2\,d\Omega.
\ee

We consider solutions for $(f_a\,/\,M_P)^2=\delta\ll 1$. When $\delta=0$ then the solution of the Einstein equations is reduced to 
$A=B=1$ (flat metric).   When $\delta$ is small, we can expand the equations of motion in a power series of $\delta$. We will retain 
only up to first order contributions in $\delta$.  Even for axions of Grand Unified Theories, where $\delta = O(10^{-4})$, second 
order contributions can be safely neglected.

The classical potential for the radial wave function, $R$, of the axion field is $W(R)=m^2\,f_a{}^2\,V(R\,/\,f_a)$ where $V(X)=1-\cos(X)$, 
but this is replaced by the expectation value of the tree level quantum potential (see Appendix \ref{Appendix1}), $V(X)=1-J_0(X)$ 
where $X=2\,\sqrt{N}\,R \,/\,f_a$. 

The expectation value of the Einstein equation and of the nonlinear equation of motion taken between states $\langle 
N-1|$ and $|N\rangle$ form a closed set of equations,
\bea\label{equations}
\frac{A'}{A^2\,r}+\frac{A-1}{A\,r^2}&=&\frac{f_a{}^2}{M_P{}^2}\left[\frac{E^2\,N\,R^2}{B\,f_a{}^2}+\frac{N\,R'^2}{A\,f_a{}^2}
+m^2\,[1-J_0(X)]\right],\nn\\
\frac{B'}{A\,B\,r}-\frac{A-1}{A\,r^2}&=&-\frac{f_a{}^2}{M_P{}^2}\left[\frac{E^2\,N\,R^2}{B\,f_a{}^2}+\frac{N\,R'^2}{A\,f_a{}^2}-
m^2\,[1-J_0(X)]\right],\nn\\
\sqrt{N}\,R''+\sqrt{N}\left(\frac{2}{r}+\frac{B'}{2\,B}-\frac{A'}{2\,A}\right)R'&+&A\left[\frac{\sqrt{N}\,E^2}{B}R-f_a \,m^2\,
J_1(X)\right]=0.
\eea

Introducing the dimension-free radial coordinate $z= r\,m$, using the rescaled wave function $X(z)= 2\,\sqrt{N} R(r)\,/\,f_a$, 
and defining $\epsilon=E\,/\,m$, $A=1+\delta\,a(z)$ and $B=1+\delta\,b(z)$, we obtain in leading order of $\delta$
\bea\label{equations2}
a'&=& -\frac{a}{z}+z\left[\frac{1}{4}\,\epsilon^2X^2+\frac{1}{4}X'^2+1-J_0(X)\right],\nn\\
b'&=&\frac{a}{z}+z\left[\frac{1}{4}\,\epsilon^2X^2+\frac{1}{4}X'^2-1+J_0(X)\right],\nn\\
X''&=&\left[-\frac{2}{z}+\frac{\delta}{2}(a'-b')\right]X'-\epsilon^2(1+\delta\,a-\delta\,b)\,X+2\,(1+\delta\,a)\,J_1(X).
\eea
$a$, $b$, and $X$  must be regular functions of $z$ at $z=0$ and vanish at $z=\infty$. While the radial wave function $X$ 
and the metric function $b$ have finite values at $z=0$, $a$ must vanish at  $z=0$.  Aside from possible initial conditions, 
the equations depend on $\delta$ and the rescaled energy, $\epsilon$.

\subsection{Expansion in the binding energy of axions}

Note that the dimension free coordinate, $z= r\,m$, measures the radial distance in units of the Compton wave length of the axion.
We will show now that for axion stars with $\delta\ll 1$ and $\Delta = \sqrt{1-\epsilon^2} \simeq \sqrt{2(m-E)/m} \ll 1$, 
(\ref{equations2}) can be reduced to a system of equations depending on $\epsilon$ and $\delta$ through the combination 
$\lambda= \sqrt{\delta}\,/\,\Delta = f_a\,/\,(M_P\,\Delta)$ only.  The parameter $\lambda$ is not in general small.   We will 
investigate bounds on possible values of $\lambda$, consistent with our approximations, later.  

To expand our equations in $\Delta$ first we expand the potential in a power series of the radial wave function, as
\be\label{pot}
V(X)=\frac{1}{4}[X^2-\frac{1}{16} X^4+\frac{1}{576} X^6+...].
\ee
Substituting (\ref{pot}) into (\ref{equations2}) the axion equation of motion takes the form
\be\label{KG}
X''=\Delta^2\,X \,[1+\delta\,(a-b)]-\left[\frac{2}{z}+\frac{\delta}{2}(b'-a')\right]X'-(1+\delta\,a)\left(\frac{1}{8}\, 
X^3-\frac{1}{192}\,X^5+...\right)+\delta\, b\, X.
\ee
It is easy to see that a systematic expansion of (\ref{KG}) in powers of $\Delta$ can be performed if we further rescale the 
dimensionless radial coordinate by introducing $x=\Delta\,z$ and simultaneously rescale the wave function as $X(z)=\Delta\, Y(x)$, 
while keeping the scale of the dimensionless metric components, $a$ and $b$, unchanged.  The powers of $\Delta$ extracted 
from each term are the ``engineering'' dimensions of  the corresponding operator.  All terms of (\ref{KG}) are of dimension 
three, except for the irrelevant, non-renormalizable terms, like $X^5$ and even higher powers of $X$, and of the relevant last 
term on the right hand side of (\ref{KG}), which is of dimension one.  Performing a systematic expansion in $\Delta$ and 
keeping relevant and marginal terms only is tantamount to taking the infrared limit of the theory.  

The leading order equations for the dimensionless field $Y(x)$, depending on the dimensionless coordinate $x$ are
\bea\label{equations3}
a'(x)&=&\frac{x}{2}\,Y(x)^2-\frac{a(x)}{x},\nn\\
b'(x)&=&\frac{a(x)}{x},\nn\\
Y''(x)&=&Y(x)-\frac{2}{x}Y'(x)-\frac{1}{8}\,
Y(x)^3+\lambda^2\,b(x)\,Y(x).
\eea
As we will see later, $\lambda=\sqrt{\delta}\,/\,\Delta$, which is the only input parameter, determines the mass and radius of 
the axion star. We will also see that $\lambda$ is a double valued function of $N$, the number of axions in the system. $N$ is 
the natural physical input parameter.

Leading order corrections to (\ref{equations3}) are of $O( \delta)$ and of $O(\delta\,\lambda^2)$, which both should 
be much less than 1.  Then in addition to $\Delta=f_a\,/\,(M_P\,\lambda) \ll 1$ we must have $\delta\,\lambda^2=(\lambda\,f_a\,/\,
M_P)^2 \ll 1$. These constraints can be combined together to give the range of validity of (\ref{equations3}) as  $f_a\,/\,M_P\ll 
\lambda\ll M_P\,/\,f_a$.  Since e.g. for QCD axions $f_a\,/\,M_P\sim 10^{-7}$, solving the system (\ref{equations3}) provides a 
correct solution for a wide range of sizes of axion stars.  

We used the shooting method to integrate (\ref{equations3}) and calculate the function, $Y(x)$. Requiring the regularity of $Y$, 
$a$, and $b$ at $x=0$ we are left with two integration constants, which can be chosen as the value of $Y(x)$ and $b(x)$ at the 
center of the star.  The boundary conditions are that $Y(x)$, $a(x)$, and $b(x)$ tend to zero at $x\to\infty$.  $a(x)$ and $b(x)$ 
have newtonian asymptotics of $a,\,b\sim x^{-1}$ at $x\to\infty$.  Such boundary conditions are difficult to implement in the numerical calculations.  However, notice that (\ref{equations3}) implies
\be
a(x)+b(x) =-\frac{1}{2}\int_x^\infty \xi \,Y(\xi)^2\,d\xi
\ee
and, consequently $a(x)+b(x)$ tends to zero exponentially when $x\to\infty$ because $Y(x)$ also has similar behavior. Such a 
boundary condition can be imposed easily in a numerical calculation.

\begin{figure}[htbp]
\begin{center}
\includegraphics[width=4in]{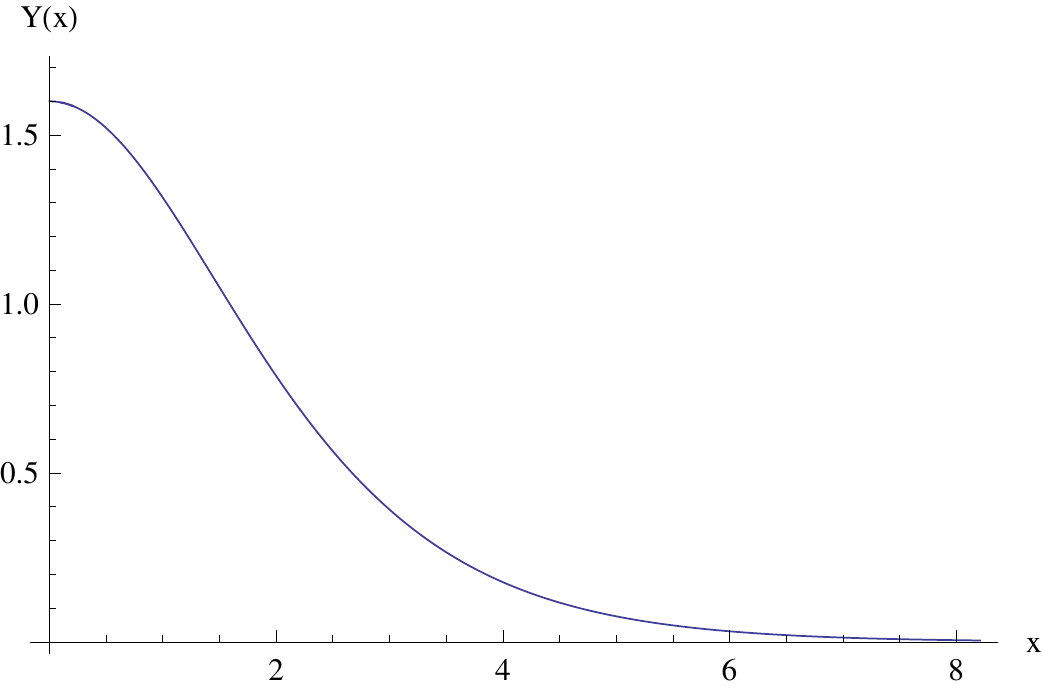}
\caption{$Y(x)$ as a function of $x$ for $\lambda=1$}
\label{Yplot}
\end{center}
\end{figure}

We performed the numerical integration of (\ref{equations3}) for a series of values of $\lambda$. As 
an example, in Fig.\ref{Yplot} we plot our solution for the wave function, $Y(x)$, as a function of $x$ at $\lambda=1$. For 
all choices of $\lambda$ we have considered, the wave function has a similar general shape.  The initial values required for  
attaining the necessary asymptotic behavior are approximately proportional to the inverse of $\lambda$.

\section{Analytic approximations to the physical parameters of axion stars}
\setcounter{equation}{0}

The most important parameters describing axion stars are the total mass, $M$, the radius inside which $99\%$ of the 
matter contained in the star is concentrated, $R_{99}$, and the number of axions in the star, $N$.   The mass of 
the star is given in leading order of the infrared limit ($\Delta\to0$) by
\be\label{mass}
M=-\frac{4\,\pi}{3}\int_0^y \,T_{00}\,r^2 \,dr\simeq\frac{f_a \,M_P}{m_a}\,\lambda\,V(\infty)=\frac{f_a{}^2}{m_a\,\Delta}\,V(\infty).
\ee
where
\be\label{Vexpr}
V(y)=\frac{2\,\pi}{3}\int_0^y \,Y(x)^2\,x^2 \,dx.
\ee

Furthermore, using the standard definition for boson stars, we define $x_{99}$ as
\be\label{x99expr}
\frac{V(x_{99})}{V(\infty)}=0.99.
\ee
Then restoring the appropriate scale, the radius of the axion star becomes 
\be\label{R99}
R_{99}=\frac{1}{m\,\Delta}\,x_{99}=\frac{M_P}{m_a\,f_a}\,\lambda\, x_{99}.
\ee
Combining (\ref{mass}) and (\ref{R99}) we obtain the relationship between its radius and its mass.
\be\label{massradiusratio}
R_{99}=\frac{1}{f_a{}^2}M\,\frac{x_{99}}{V(\infty)}.
\ee

Using the function $Y(x)$ found by numerical integration we calculated $V(\infty)$ and $x_{99}$ for a series of values 
$0.01 \leq \lambda\leq 10$.  As they are functionals of $Y(x)$, they are uniquely fixed by the value of $\lambda$.  The ratio 
$x_{99}\,/\,V(\infty)$, appearing in (\ref{massradiusratio}) is larger than 0.1 throughout the range of $\lambda$ we 
consider, leading to the relationship
\be\label{massradiusratio2}
R_{99}\gtrsim \frac{0.1}{\delta}\frac{M}{M_P{}^2}\simeq \frac{2\times 10^{-3}}{\delta}R_{\rm S},
\ee
where $R_{\rm S}$ is the Schwarzschild radius.  This relationship, which implies gravitational stability is satisfied 
provided $\delta<< 1\,/\,500$.   As shown by (\ref{massradiusratio2}), our formalism could even be applied to axions 
with  $f_a\sim 10^{15}-10^{16}$ GeV, because  for axions with decay constant $f_a \approx 10^{16}$ GeV, $\delta \sim 10^{-4}$. 

At $\lambda<<1$ a  linear fit gives an excellent representation of $R_{99}$,
\be\label{ratsmalllambda}
R_{99}(\lambda)\simeq \frac{M_P}{f_a\,m_a}\,2.735\,\lambda=\frac{1}{\Delta\,m}\,2.735.
\ee
For $\lambda\gtrsim0.5$ the rise of $R_{99}$ is steeper,
\be
R_{99}(\lambda)\simeq \frac{M_P}{f_a\,m_a}\,(0.456+5.75\,\lambda)
\ee
gives an excellent fit.  $R_{99}$ continues to increase throughout the range of $\lambda$ considered. The axion stars 
investigated in \cite{Barranco} correspond to the extremely small values of $\lambda \sim 10^{-6}-10^{-5}$.

Using fit (\ref{ratsmalllambda}),  we obtain  from (\ref{massradiusratio}) at $\lambda<<1$
\be
M(\lambda)\simeq \frac{M_P\,f_a}{m_a}\,50.26\,\lambda=50.26 \frac{f_a{}^2}{m_a\,\Delta}.
\ee
 \begin{figure}[htbp]
\begin{center}
\includegraphics[width=4in]{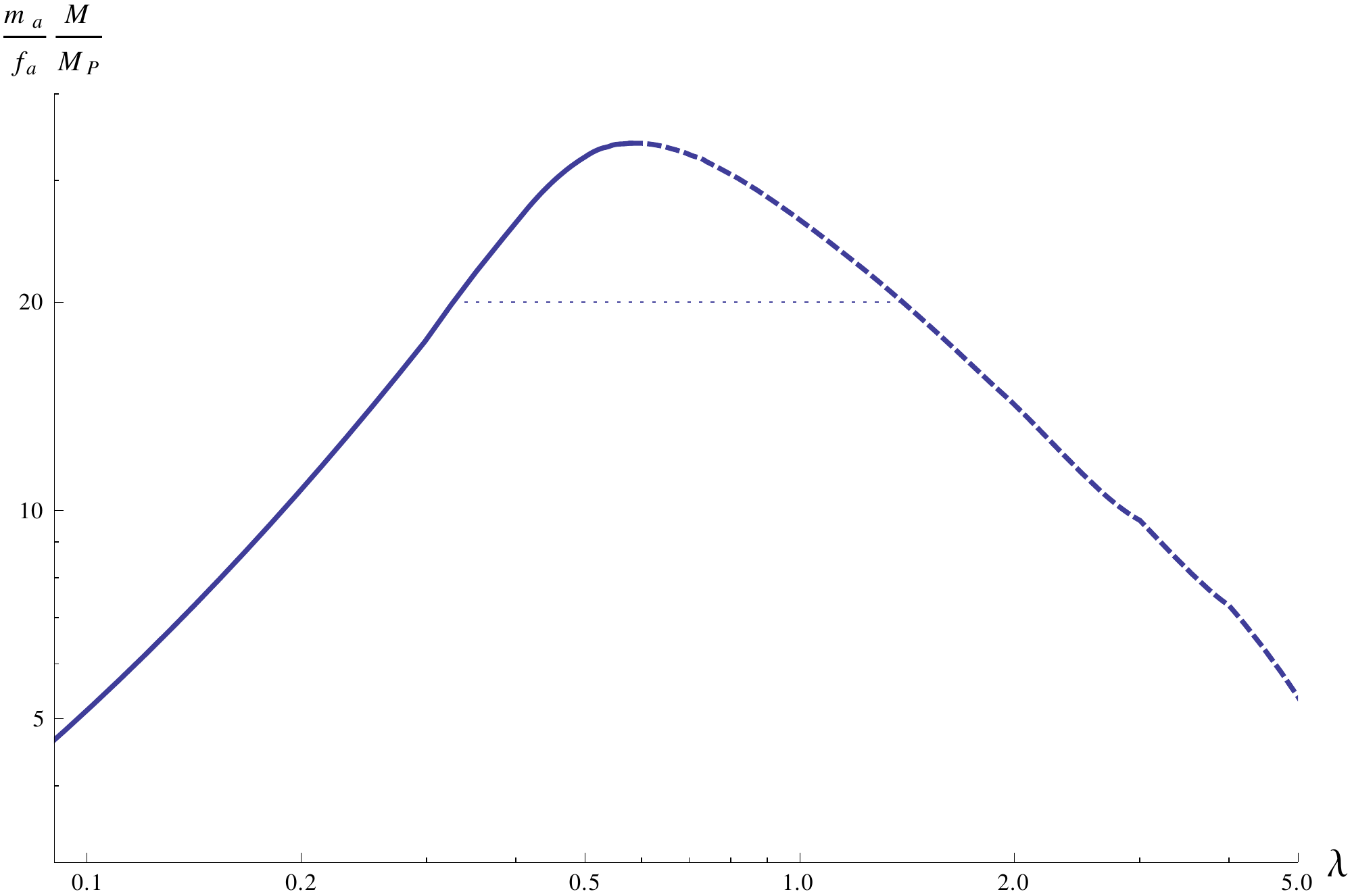}
\caption{The mass $M$, scaled by the factor $m_a\,/\,(f_a\,M_P)$ as a function of $\lambda$. Stable (unstable) stars 
are represented by the solid (dashed) line. The dotted line connects sample states with identical numbers of axions. The 
mass difference of these stars is a small fraction of the mass.}
\label{Mplot}
\end{center}
\end{figure}
However, at  $\lambda\gtrsim1$ (very weak binding) $V(\infty)\simeq 15\,\lambda^{-2}$, consequently (\ref{mass}) gives
\be\label{mass2}
M=\frac{f_a{}^2}{m_a\,\Delta}\,V(\infty)\simeq 15\,\frac{M_P{}^2}{m_a}\Delta.
\ee

We plotted the dependence of the mass, scaled by the factor $f_a\,m_a\,/\,M_P$, as a function of $\lambda$ in Fig. 
\ref{Mplot}.  The striking feature of the plot is the maximum of the mass as a function of parameter 
$\lambda=f_a\,/\,(M_P\,\Delta)$.  Maxima of $M$, as a function of the central value of the axion wave function have 
already been found in previous work \cite{CSW}, \cite{bosonstars}, \cite{Barranco}, but the maximum in Fig.\ref{Mplot} 
has a physical significance  due to the following considerations.\footnote{Seidel and Suen ~\cite{SS} have discussed 
the question of instability  in a complex scalar field model of boson stars.}  Note that the number of axions in 
the condensate is approximately equal to
\be\label{number}
N=\frac{M(\Delta)}{m_a\,\sqrt{1-\Delta^2}}.
\ee  
The maximum, $M_{\rm max}=M(\Delta_{\rm max})$ is attained at $\lambda_{max}= \sqrt{\delta}\,/\,\Delta_{max}\simeq0.58.$  
Then the maximal number of axions in an axion star is 
\be\label{NandM}
N_{\rm max}\simeq\frac{M_{\rm max}}{m_a\,\sqrt{1-\Delta_{\rm max}{}^2}}.
\ee  
For every $N<N_{\rm max}$, there are two masses, $M(\Delta_1)$ and $M(\Delta_2)$, such that (\ref{number}) is satisfied. 
Assume $\Delta_1>\Delta_2$, or equivalently, $\lambda_1<\lambda_{\rm max}<\lambda_2$.  Then $M(\Delta_1)<M(\Delta_2)$, 
hence the relationship between the masses is illustrated by the dotted line in Fig. \ref{Mplot}. As a result, every 
star corresponding to the branch of the curve at  $\lambda>\lambda_{max}$, which is plotted by a dashed line, is unstable. 
In fact, using (\ref{NandM}) it is easy to  estimate the mass difference between the two states of the axion star containing 
$N$ axions. We obtain
\be\label{massdifference}
\frac{\delta M}{M}\simeq\frac{1}{2}(\lambda_1{}^{-2}-\lambda_2{}^{-2})\,\delta .
\ee
$\delta M\,/\,M<<1$, but as is shown in Table \ref{tab1},  $\delta M$, still represents a substantial amount of macroscopic energy.

Now, admittedly, there is an energy barrier between states $M(\Delta_2)$ and $M(\Delta_1)$. In the present paper we do 
not attempt to calculate the decay rate.  That will be the subject of future research.  

\begin{table}
\caption{Mass, radius, average density, and $\delta M$, as defined in (\ref{massdifference}),  as a function of $\lambda=f_a \,/\,(M_P\,\Delta)$}
\centering
\vskip 0.2 cm
\begin{tabular}{| >{$}l<{$} | >{$}l<{$} | >{$}l<{$} |>{$}l<{$}  |>{$}l<{$}|}
\hline
\lambda  & M\,\text{(kg)}  &  R_{99}\,\text{(km)}&\text{d\,
( kg\,/\,m}^3 )&  \delta M\,(\text{kg}) \\
   \hline
 0.1 & 1.34\times 10^{18} & 115 & 207 & 167000. \\
 0.3 & 4.61\times 10^{18} & 386 & 19.1 & 61700. \\
 0.4 & 6.78\times 10^{18} & 593 & 7.74 & 44700. \\
 0.5 & 8.44\times 10^{18} & 854 & 3.24 & 21700. \\
 0.54 & 8.74\times 10^{18} & 972 & 2.27 & 11100. \\
 0.58 & 8.84\times 10^{18} & 1076 & 1.69 & 1570. \\
 0.62 & 8.81\times 10^{18} & 1183 & 1.27 & -7160. \\
 0.8 & 7.98\times 10^{18} & 1652 & 0.422 & -30900. \\
 1 & 6.85\times 10^{18} & 2145 & 0.166 & -44100. \\
 2 & 3.71\times 10^{18} & 4499 & 0.0097 & -71200. \\
 4 & 1.9\times 10^{18} & 9062 & 0.0006 & -11800. \\
 10 & 7.65\times 10^{17} & 22849 & 0.000015 & -355000.
\\
  \hline
\end{tabular}
\label{tab1}
\end{table}

In Table \ref{tab1} we provide the mass,  mass difference between stars containing the same number of axions ($\delta M$), 
radius, and density of axion stars for several values  of $\lambda$ between $10^{-1}\leq 
\lambda\leq 10$.  (Note that this range is well inside the interval $f_a/M_P\ll\lambda\ll M_P/f_a$, where our 
approximations are valid.)  For the input values of $f_a$ and $m_a$ we use  QCD axions, with parameters satisfying  $m_a = 6\,\mu 
{\rm eV} \times 10^{12}\,{\rm GeV}\,/\,f_a$ \cite{PS} \cite{CA}.  If we pick the value $m_a = 10^{-5}$ eV, we obtain $f_a=6  
\times 10^{11}$ GeV.  Note, however, that  in Table \ref{tab1}, at fixed $m_a\,f_a$, the values of  $M$,   $\delta M$ and $d$ 
are increasing functions of $f_a$, while $R_{99}$ is independent of the choice of $f_a$.

A final comment concerns the critical temperature of the axion condensate. To see that the axion star is a pure condensate, 
with very little contribution from excited states one needs to consider its critical temperature.  Using  the average density, 
$\rho=M\,/\,(m_a\,V)\simeq 3\,M\,/\,(4\,\pi \,R_{99}^3\,m_a)$, and the expression for the radius in (\ref{massradiusratio2}) 
we obtain the following estimate for the critical temperature
(neglecting interactions)
\be\label{critical}
T_c=\frac{2\,\pi}{m_a}\left(\frac{\rho}{2.612}\right)^{2/3}\simeq 12.7\, m_a \left(\frac{f_a{}^4}{m_a^2\,M_P{}^2}\right)^{2/3}.
\ee
For a QCD axion, with $f_a \approx 6 \times 10^{11}$ GeV, this would give $T_c\sim 10^{11}$ GeV.

\section{Conclusions}

As it has been pointed out in the previous section, to describe compact  objects formed from axions one needs to consider that, 
except possibly in the early universe, the temperature of the object is many orders magnitudes smaller than the critical 
temperature of boson condensation. In other words, in a good approximation one may assume that all the axions are in the ground 
state. Therefore, along with Ruffini and Bonazzola \cite{RB} we substitute  the wave function of the axions by a quantized field.  
As the resulting complicated nonlinear field theory cannot be solved analytically, one is restricted to use perturbation theory. 
A similar approach was used by Barranco and Bernal \cite{Barranco} in their application to axion stars, but they explore a region 
in which parameter $\lambda$ is extremely small, $\lambda =O(10^{-6})$.

The most important observations of this paper are that the application of the method of \cite{RB} to axion stars must be restricted 
in two different ways: (i) the decay constant of the axions, $f_a$ must be much smaller than the Planck mass, or using the notations 
of this paper $\delta=f_a{}^2\,/\,M_P{}^2\,<<\,1$; (ii) only weakly bound axions should be considered, i.e. the axion binding 
energy should satisfy $m_a -E \ll m_a$. The reason for requiring small binding energy is a requirement of using the Ruffini-Bonazzola 
method \cite{RB} for an interacting field theory:  if the binding energy is of $O(m_a)$ then the perturbation expansion of the axion 
field theory breaks down. Pairs of positive and negative energy states contribute  to the ground state and to the expectation values 
of physical quantities, as well.  To extend calculations beyond the region $\delta\ll1$ and $\Delta =\sqrt{1-E^2\,/\,m_a^2}\ll1$ 
requires using non-perturbative field theory, which is beyond the scope of this paper. 

Requiring $\Delta\ll 1$ has allowed us to expand the equations of motion in the scale parameter $\Delta$.  If we consider 
that the scaling dimension of the axion field is unity, the expansion in $\Delta$ is tantamount to taking the infrared limit of the 
axion field theory.  Only marginal terms and a relevant term with coefficient $\lambda^2=\delta\,/\,\Delta^2$  of the Lagrangian give 
a contribution in leading order of $\Delta$.  Consequently, sixth and higher order terms of the transcendental axion potential do not 
contribute.   

We have calculated the masses, radii and densities of axion stars as a function of the parameter $\lambda=\sqrt{\delta}\,/\,\Delta$ 
over two orders of magnitude of the variable $\lambda$, consistent with the restrictions $\delta,\,\Delta\ll 1$. We found that the 
radius of the star increases approximately linearly throughout the range of $\lambda$ we considered. However, while the mass of the 
star, $M(\lambda)$, rises linearly at small $\lambda$, it reaches a maximum at $\lambda_{\rm max}\simeq 0.58$ and tends to zero at 
large $\lambda$. The number of axions in the system, $N(\lambda)$, which is determined uniquely by $\lambda$, is the same at two 
different values, $\lambda_1$ and $\lambda_2$, where $\lambda_1<\lambda_{\rm max}<\lambda_2$.  We show (see eq. 
(\ref{massdifference})) that $M(\lambda_2)>M(\lambda_1)$.  This implies that all states on the branch $\lambda>\lambda_{\rm max}$ 
of the $M(\lambda)$ curve  are unstable. 

For a QCD axion, assuming $m_a=10^{-5}$ eV  we obtained $M_{\rm max}\equiv M(\lambda_{\rm max})\simeq  10^{19}$ kg  and 
$R_{\rm max}=R(\lambda_{\rm max})\simeq1000$ km.\footnote{Note that $R_{\rm max}$ is the radius of the heaviest  axion star but 
not the axion star of largest radius.  The radii of unstable axion stars are always larger than $R_{\rm max}$.} For fixed 
$f_a\,m_a= 6\times 10^{-3}\,{\rm GeV}^2$ the radius is independent of $f_a$, while the mass is an increasing function of  $f_a$. The maximal 
number of axions in an axion star is $N_{\rm max}=7.3\times 10^{58}$, but it is a fast increasing function of $f_a$. 

In the future we will also investigate rotating axion stars and analyze the maximal mass and the question of stability as a 
function of angular momentum.  We will investigate collisions of axion stars in view of the existence of the maximal mass. 
Furthermore if an axion star is in isolation, unstable states, with $\lambda>\lambda_{\rm max}$ should decay into the 
corresponding stable state.  We will compute  rate of transition, and possible signatures of their decay.

The masses of compact axion star solutions found in our work are consistent with the mass bounds derived by Tkachev for condensate 
formation through gravity \cite{tka91}. Axion stars, along with free axions may form all or part of dark matter.  Various production 
mechanisms of axions in the early universe are discussed in \cite{pww83}. Bounds on  the axion decay constant resulting from 
astrophysical and cosmological observations are discussed in \cite{{pww83}} and \cite{don69}. We will investigate how our conclusions 
change if axion stars are in equilibrium within a cloud of free axions after reaching the maximal mass. We will also investigate 
the consequence if axions come in a multitude of flavors, as expected in theories derived from compact extra dimensions. 

The collisions of axion condensates with neutron stars have been studied in the past \cite{Barranco2}, \cite{iwa99}. Recently axion stars have been 
proposed as progenitors for Fast Radio Bursts (FRBs) \cite{lb07} via collisions with neutron star atmospheres \cite{it10}. 
Whatever produces FRBs should be able to generate large amounts of energy, on the order of approximately $10^{42}$ ergs/s, in a 
fairly tight frequency range around 1.4 GHz over time scales on the order of milliseconds.  Assuming that all of the mass of an 
axion star is converted into radiation during a putative collision with a neutron star, one finds that an axion star mass on the 
order of $10^{-12}~ M_\odot$ would be required. This mass is indeed compatible with our results listed in Table 1. However, as 
the radius of our boson star is at least an order of magnitude larger that that of a neutron star we feel that the possibility of converting 
all the axions to radiation in such a collision is highly unlikely. Therefore we plan to perform an accurate estimation of this 
conversion rate in the near future. 

The radius of the axion stars may be estimated from the duration of the bursts and was found in \cite{lb07} to be on the order of 
100 km, which is also compatible with our results in Table 1. Moreover, \cite{lb07} also showed that the frequency of the radiation 
can be generated by axions of mass approximately $10^{-5}$ eV. We hope to examine this mechanism in greater detail and will report 
on the results of our investigation elsewhere.

As mentioned, one result of our calculations is that for a fixed number of axions there are in general two masses corresponding 
to two possible values of $\Delta$. An intriguing consequence is the possibility of tunneling from the state with a higher mass 
to the one with a lower mass. We will investigate whether collisions with neutron stars could induce such a tunneling process 
resulting in a fainter companion burst of approximately $10^{29}$ ergs, where we have taken $\delta M \sim 10^5$ kg for an axion 
star of mass $10^{18}$ kg and radius 100 km. If detected, such a companion burst could serve to distinguish between axion star 
progenitors of FRBs and other proposals \cite{tot13}. However, considering the faintness of the companion burst, it is unlikely 
to be observed from events that occur outside our galaxy. The observed frequency of FRB events (about $10^{-3}$ events per galaxy 
per year) would therefore make this phenomenon even more difficult to observe.

Finally, we would like to emphasize again that there is no fundamental reason to limit  calculations to theories with decay 
constants satisfying $f_a\ll M_P$ and stars with sufficiently small binding energy. We just state that the method of taking the 
expectation value of the equations of motion in free particle states, which are defined in flat space is not permissible if the 
axion decay constant is comparable with the Planck mass and/or the relative binding energy is not much smaller than 1. Using the 
current formalism to extend  the results presented in this paper beyond leading order expansion in $\delta$ and $\Delta$ would 
lead to false results.  Such an extension would require using exact solutions of an interacting field theory, possibly in curved 
space, a calculation, which is beyond the scope of our current analysis.  Restricting ourselves to leading order 
contributions has the added benefit of simplifying results to the extent of being able to obtain  a clearer physical interpretation 
of the properties of axion stars.

\section*{Acknowledgements}
We thank P. Argyres, P.Esposito, A. Kagan, C. Kouvaris, and J. Zupan for discussions.

\appendix 

\section{Calculation of the quantum potential in tree approximation}\label{Appendix1}
\setcounter{equation}{0}

Following \cite{Barranco}, field $\Phi$ can be promoted to a scalar quantum field  as
\be\label{phi}
\Phi(x)=\Phi^-(x)+\Phi^+(x)\equiv\sum_{n,l,m}R_{n,l}(r)\,[Y_l^m(\theta,\phi)\,a_{n,l,m}+Y_l^{m\,\star}(\theta,\phi)\,a_{n,l,m}^\dagger],
\ee
where $a_{n,l,m}$ is the annihilation operator of a state with radial quantum number $n$ and angular quantum numbers $l$ and $m$.   
Note that the commutator $[\Phi^+,\Phi^-]$ is a c-number. In particular
\be
[a_{n,l,m}^\dagger \,,a_{n',l',m'}]=\delta_{n\,n'}\delta_{l\,l'}\delta_{m\,m'}
\ee
with the rest of the commutators vanishing.

Let us consider now the factor $ 1-\cos({\cal X})$ of the axion potential, where the operator ${\cal X}=\Phi(x)\,/\,f_a$.  If we omit 
loop contributions the expectation value of the cosine in an $N$-particle condensate can be calculated exactly, without resorting to 
the Taylor expansion. The Baker-Campbell-Hausdorff lemma implies that, if $[{\cal X}^+[{\cal X}^+,{\cal X}^-]=0$, then 
\be\label{BCH}
\cos({\cal X}^+\,+{\cal X}^-)=e^{-\frac{1}{2}[{\cal X}^+\,,{\cal X}^-]}\,e^{i\,{\cal X}^+}\,e^{i\,{\cal X}^-}+c.c.
\ee
Taking the expectation value of the operator of (\ref{BCH}) between $N$ particle condensates using (\ref{phi}) 
one obtains\footnote{Note that the definition of the state is $\vert N\rangle=(a^\dagger)^N\vert 0\rangle\, (N!)^{-1/2}$, where 
$a^\dagger$ is the creation operator of a ground state particle.}
\be\label{cosine}
\langle N \vert \cos[(\Phi^+\,+\Phi^-)\,/\,f_a]\vert N\rangle=e^{-\frac{1}{2\,f_a{}^2}[\Phi^+\,,\Phi^-]}\langle N\vert  
e^{i\,\Phi^+\,/\, f_a}\,e^{i\,\Phi^-\,/\,f_a}\vert N\rangle+c.c
\ee
Omitting loop corrections and using free particle states, the right hand side of (\ref{cosine}) can be readily calculated after 
expanding the two exponentials into power series of the creation and annihilation operators.  Owing to the fact that in the 
condensate only ground state particles can be annihilated in the expectation values all contributions containing operators of 
the excited states vanish on the right hand side of (\ref{cosine}). It follows that
\be\label{sum}
\langle N\vert  e^{i\,\frac{\Phi^+}{f_a}}\,e^{i\,\frac{\Phi^-}{f_a}}\vert N\rangle=\sum_{k=0}^N\frac{(i\,\frac{ R}{f_a})^{2\,k}}
{(k!)^2}\frac{N!}{(N-k)!},
\ee
where $R=R_{1,0,0}$ is the single particle ground state radial wave function.

When $N\to\infty$ at fixed $R\,/\,f_a$, the sum is dominated by $k<<N$. Then the last multiplier on the right hand side of 
(\ref{sum}) tends to $N^k$ giving our final result
\be
\langle N \vert \cos(\Phi\,/\,f_a)\vert N\rangle=e^{-S}\,J_0(2\,\sqrt{N}\,R\,/\,f_a),
\ee
where 
\be
S=\frac{1}{2\,f_a{}^2}\sum_{n,l,m}R_{n,l,m}^2|Y_{l,m}|^2.
\ee
If we introduce the rescaled field, $X(z)=2\,\sqrt{N}\,R\,/\,f_a$ the exponent $S$ acquires a factor $N$ in the denominator, while the 
$N$-dependence cancels in all other terms of the equations of motion. Thus, the in the limit of $N\to\infty$ we can set $S\to 0$ leaving 
the tree level expectation value of the axion potential $V_q(X)=m^2\,f_a{}^2\,[1-J_0(X)]$. Contrast this with the classical potential, 
$V_c(X)=m^2\,f_a{}^2\,[1-\cos(X)]$. Note that the expansion terms are different for the two potentials. For small $X$ (i.e. for large 
distances), where quadratic contributions dominate, $V_c(X)\,/\, V_q(X)\simeq 2$.

When one calculates the expectation value of the scalar equation in the semiclassical approximation one should evaluate
\be
\langle N-1\vert \Box\Phi -\frac{1}{2}V'(\Phi\,/\,f_a)\vert N\rangle=0,
\ee
which equation turns into differential equation (\ref{equations2}), if one sets $\Phi=f_a\,X (a^\dagger\,e^{i\,E\,t}+a \,e^{-i\,E\,t})$, 
with $V'(X)=J_1(X)$.

\end{document}